# DESORPTION OF EXCITED MOLECULES FROM SOLID NITROGEN

Elena Savchenko [*,1], Ivan Khyzhniy[1], Sergey Uyutnov[1], Andrey Barabashov[1],
Galina Gumenchuk[2], Alexey Ponomaryov[3], and Vladimir Bondybey[2]

[1] Institute for Low Temperature Physics and Engineering NASU, Lenin Ave. 47, 61103 Kharkov, Ukraine
[2] Lehrstühl für Physikalische Chemie II TUM, 85747 Garching, Germany
[3] Helmholtz Zentrum Dresden-Rossendorf, 01328 Dresden, Germany

The role of charged centers in radiation-induced phenomena occurring in solid nitrogen irradiated with an electron beam was studied. The experiments were performed employing luminescence method and activation spectroscopy techniques – spectrally resolved thermally stimulated luminescence TSL and thermally stimulated exoelectron emission. To discriminate radiation-induced processes in the bulk and at the surface the samples were probed in depth by varying electron energy. Desorption of excited $N_2^*$ ($C^3\Pi_u$) molecule was detected for the first time. The mechanism of this phenomenon based on recombination of electron with intrinsic charged center $N_4^+$ was proposed. The key role of $N_3^+$ center dissociative recombination in generation of N radicals is suggested.

## 1 Introduction

Radiation effects in solid $N_2$ attract much attention in diverse fields such as material and surface sciences, physics and chemistry of interstellar and solar systems, particle physics. Desorption or sputtering is among the most intensively studied radiation-induced phenomena. Electronic desorption of solid nitrogen was studied under excitation with electrons [1-5], ions [6-12] and photons [13, 14]. Current state of the problem is given in [15]. Despite extensive studies mechanisms of electronic sputtering and the part of charged defect centers are still not well understood. The experiments [16] have revealed thermally stimulated exoelectron emission (TSEE) from pre-irradiated α-phase $N_2$ films which indicates formation and accumulation of charged centers. It is worthy of note that electrons in the α-phase $N_2$ are highly mobile [17]. Their localization at the lattice imperfections and impurities with positive electron affinity produces negatively charged defect centers. Probability of $(N_2)_2^+$ centers formation was mentioned in [18] while discussing the energy distribution of particles desorbed from solid $N_2$ by electron impact. The same centers were suggested to contribute to the novel phenomenon – anomalous low-temperature post- desorption ALTpD observed upon heating of a preliminary irradiated solid $N_2$ [19]. Creation of ionic species $N_3^+$ in electron-bombarded solid $N_2$ was reported recently [20].

In the present work, we applied luminescence spectroscopy and activation spectroscopy methods in order to get information on charged defects and elucidate their role in defect-induced processes in electron-bombarded films of solid $N_2$. Recording the luminescence spectra sequences on exposure time enabled us to monitor defect production, excited particle ejection and fragmentation of molecules. Surface- and bulk-related centers were discriminated by varying electron energy, i.e. the penetration depth. Relaxation processes were monitored using a concurrent measurement of three

---
* Corresponding author: e-mail elena.savchenko@gmail.com, Phone: +38 057 341 0836, Fax: +38 057 340 3370

relaxation emission – photons, electrons and nitrogen particles. The obtained information on „post-irradiation" processes provides new insights into scenario and relaxation paths that are realized under irradiation of solid $N_2$ and clarifies the role of charged defect centers.

## 2 Experimental section

The experiments presented here were performed using facilities at TUM and ILT described in more detail in (Ch. 7 by E. V. Savchenko in [21]). In addition to luminescent spectroscopy we used activation spectroscopy methods because detailed knowledge of the processes occurring after completing irradiation can provide new insights into the radiation effects and into the stability and dynamics of both the charged carriers and the neutral transient species. As it was demonstrated in [22] the use of current activation spectroscopy methods along with traditional thermally stimulated luminescence (TSL) is essential to distinguish reactions of charged and neutral centers. Taking into account the observation of TSEE from pre-irradiated solid $N_2$ [16] we employed concurrent measurement of TSEE, spectrally resolved TSL in the visible and VUV ranges and yield of desorption by pressure monitoring. TSEE current was detected with an electrode kept at a small positive potential +9 V and connected to the current amplifier. When measuring the relaxation emissions we used heating at a constant rate of 5 K/min. The entire control of the experiment and the simultaneous acquisition of the TSL and TSEE yields, as well as measuring and recording the sample temperature and pressure were accomplished using a computer program developed specifically for these studies.

The samples were grown from the gas phase by deposition on a metal substrate coated by a thin layer of $MgF_2$, which was cooled to about 7 K by a two stage, closed-cycle Leybold RGD 580 refrigerator or a liquid helium cryostat. High-purity (99.999%) $N_2$ gas was used. The base pressure in the vacuum chamber was about $10^{-8}$ mbar. The deposition rate and the sample thickness were determined by observing the pressure decrease in a known volume of the gas-handling system. The typical deposition rate was about $10^{-1}$ μms$^{-1}$, and samples of thickness 100 μm were grown.

In order to ionize the samples and produce charge centers we used slow electrons of 500 eV energy. The deposition was performed with a concurrent irradiation by electrons to generate charge centers throughout the sample or alternatively the sample was irradiated after deposition. The current density was kept at 30 μAcm$^{-2}$. An electrostatic lens was used to focus the electrons. The radiation dose was varied by an exposure time. In the experiments presented we measured yields of TSEE from the samples, TSL in VUV range and pressure above the sample during sample heating. The exoelectron yield was measured with an electrode kept at a small positive potential +9 V and connected to the current amplifier FEMTO DLPCA 200. The VUV cathodoluminescence spectra in the range 50-300 nm were recorded with a modified VMR-2 monochromator. The pressure changes in the sample chamber during the experiment were monitored using a Compact BA Pressure Gauge PBR 260. For calibration we used a flow rate controller.

The temperature was measured by a calibrated silicon diode sensor mounted directly on the substrate, with the programmable temperature controller LTC 60 allowing us not only to measure and maintain any desired temperature during sample deposition, annealing and irradiation, but also to control the heating regime flexibly. The relaxation processes in $N_2$ samples were studied in the temperature range from 7 to 45 K. The entire control of the experiment and the simultaneous acquisition of the TSL and TSEE yields, as well as measurement and recording of the sample temperature and of the vacuum sample chamber pressure were accomplished using a computer program developed specifically for these studies.

### 3 Results and discussion

In view of a wide band gap of solid $N_2$ (15.6 eV) we performed measurement of luminescence spectra over a broad wavelength range – from the visible to the VUV region. In the luminescence spectra measured we observed the well-known atomic nitrogen band, so-called α-group, related to the $^2D \rightarrow ^4S$ transition [23] and molecular series located in VUV and near UV ranges. The typical luminescence spectrum of solid nitrogen in VUV range is shown in Fig. 1.

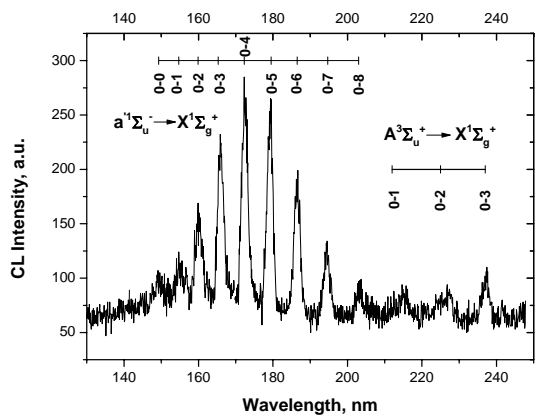

**Figure 1** Luminescence of solid nitrogen in VUV range.

It consists of the singlet progression $a'^1\Sigma_u^- \rightarrow X^1\Sigma_g^+$ and intercombination one $A^3\Sigma_u^+ \rightarrow X^1\Sigma_g^+$. Energy positions of the vibrational bands of both progressions are in good agreement with those observed in the early study of VUV luminescence of solid $N_2$ [24, 25]. Note that both progressions belong to emission of bulk excitations as is seen from a comparison of spectra detected at low and high electron beam energy. Both progressions are shifted toward lower energy with respect to the gas phase spectra. The shift for $A^3\Sigma_u^+ \rightarrow X^1\Sigma_g^+$ progression is 450 cm$^{-1}$ and 330 cm$^{-1}$ for the $A^3\Sigma_u^+ \rightarrow X^1\Sigma_g^+$ one.

The spectrum detected in the near UV and visible ranges is presented in Fig. 2. The spectrum is dominated by the α-group corresponding to the emission of N atoms initiated by matrix phonons. The α'-line represents the simultaneous vibrational excitation of $N_2$ molecule ($v=0 \to v=1$) with the atomic $^2D \to {}^4S$ transition [23]. The β-group stems from oxygen impurity and corresponds to the $O(^1S) \to O(^3P)$ transition of O atom.

In the near UV range we registered the emission of second positive system – the transitions between $C^3\Pi_u$ and $B^3\Pi_g$ excited molecular states (denoted in Fig. 2 by 2PB).

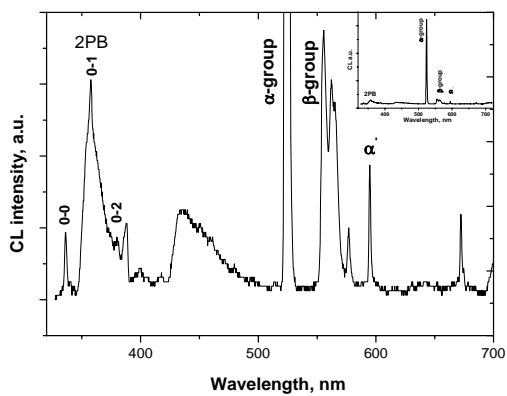

**Figure 2** Luminescence of solid $N_2$ in visible and near UV ranges excited with 1 keV electron beam. The overall view of spectra is shown in the inset.

This emission was detected previously in [26] against a background of unknown origin. The distinctive feature of this emission is the coincidence of the observed bands with those detected in the gas phase spectra within the accuracy of our measurements. Fig. 3 shows in more detail the luminescence of solid $N_2$ in the range of second positive system recorded at 5 K.

"Hot" luminescence – transitions from several vibrational levels ($v'=0, 1, 2$) of the $C^3\Pi_u$ state have been detected. Shape of the bands is difficult to analyse because of their overlap. The only band, free from overlap, is the band corresponding 0-0 transition. Its shape is characterized by strong narrow band coinciding with 0-0 term of the $C^3\Pi_u \to B^3\Pi_g$ progression in the gas phase spectrum excited by an electron beam [27]. The blue sub-band is close to the rotational R branch. Note that the blue sub-band is also close to the $(N_2)_2$ emission observed in the high-pressure discharge [28]. At present it is difficult to unambiguously assign the blue sub-band observed in our experiments. We detected also the red sub-band which was not distinguished in [26] because of overlap with some unidentified impurity. The red sub-band grows in intensity with the electron beam energy increase suggesting its bulk origin. In contrast strong narrow band coinciding with the gas band is thought to be the emission of nitrogen molecules desorbing in the excited $C^3\Pi_u$ state.

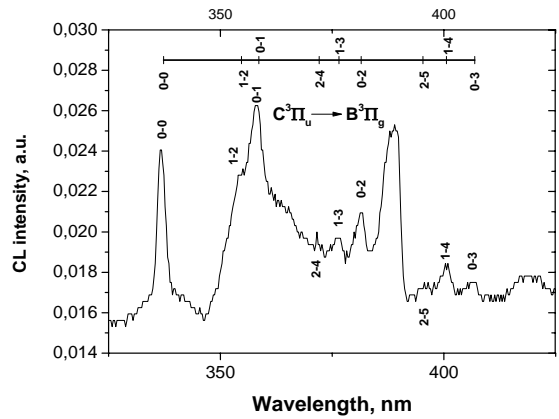

**Figure 3** The second positive system in luminescence of solid $N_2$ excited with 500 eV electron beam.

Additional experiments on probing the samples in depth by the electron beam energy variation and experiments with thin films (< 100 nm) support this assignment. Fig. 4 demonstrates the spectrum transformation with changing the penetration depth. Simple estimation and the data reported in [29] show that the penetration depth differs by an order of magnitude (from 10 nm to 100 nm) when the electron energy beam changes from 500 eV to 1,2 keV. For clarity the spectra were normalized to the intensity of α-band related to the emission of bulk centres. An increase of the ratio of the intensity of the second positive system to the intensity of α-band at lower electron beam energy is clearly seen.

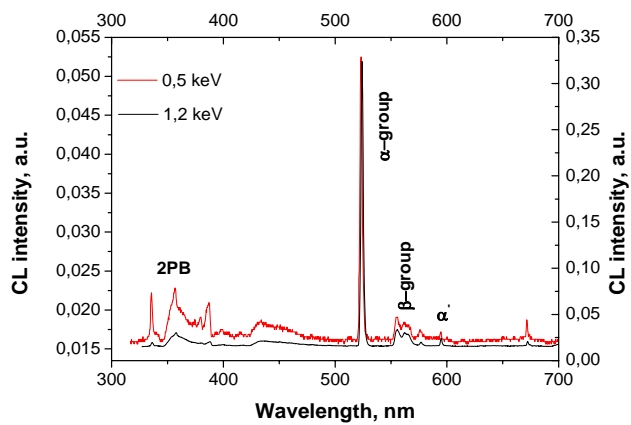

**Figure 4** Luminescence spectra of solid $N_2$ excited with 500 eV and 1 keV electron beam. For simplicity sake the spectra are normalized to the α-band intensity.

Similar effect of the molecular emission enhancement was observed in thin films as illustrated in Fig. 5.

An increase of the relative intensity in favour of the molecular emission in thin film and at lower electron beam energy supports its relation to the surface. These findings are in contradiction with the interpretation of this emission as emission of molecules freely rotating in solid nitrogen proposed in [26]. It is known that in the α-phase of solid $N_2$ rotation of molecules is frozen and the barrier to rotation is about 35 K [30]. Moreover, we observed a blue matrix shift of the bands in Ne matrix in accordance with [31]. Taken together our observations argue for the assignment of the $C^3\Pi_u \to B^3\Pi_g$ progression in luminescence spectrum as the emission from $N_2$ molecule desorbing in the excited state. The partial yield of the excited molecules desorption (monitored by the luminescence) correlated with the total desorption yield (monitored by the pressure). Note that desorption of metastable nitrogen molecules stimulated by low-energy electrons was detected by time-of-flight technique in [5].

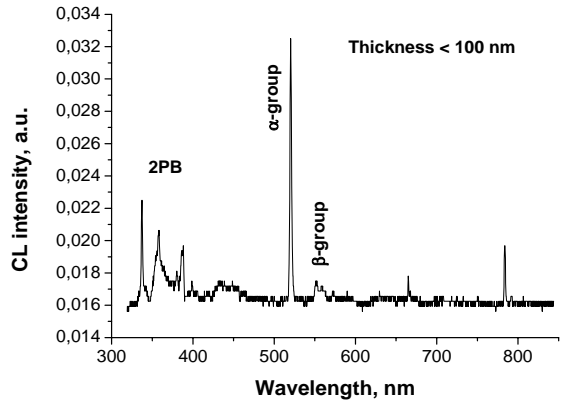

**Figure 5** Luminescence spectra of thin $N_2$ film (<100 nm).

Monitoring the luminescence spectra with an exposure time revealed accumulation of radiation-induced centers responsible for the emissions observed. Fig. 6 shows the dose dependence of the $C^3\Pi_u \to B^3\Pi_g$ transition by the example of 0-1 band. Most likely it is the result of positively charged centers $N_4^+$ formation and accumulation. Reaction of dimerization is well-known for a number of materials, e.g. rare-gas solids [32] and dense rare gases [33]. Formation of the molecular cluster $N_4^+$ in the gas phase at high pressure by the reaction:

$$N_2^+ + 2N_2 \to N_4^+ + N_2 \qquad (1)$$

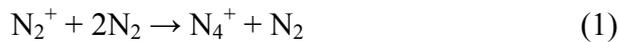

was observed in [34]. The experiments with free supersonic jet of nitrogen below 20 K have demonstrated efficiency of ionic association reaction (1), which is characterized by the inverse temperature rate law [35].

Recombination of $N_4^+$ with electron proceeds by the reaction [36]:

$$N_4^+ + e^- \to N_2^* + N_2 + \Delta E_2 \qquad (2)$$

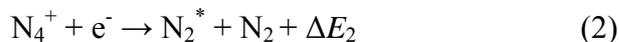

So, dissociative recombination of $N_4^+$ with electron results in the appearance of excited nitrogen molecule. Emission of the second positive system following reaction (2) was first detected in [37] using pulse radiolysis of $N_2$ at high pressure. The REMPI experiment [38] performed at the near-atmospheric pressure $N_2$ also demonstrated population of the $C^3\Pi_u$ state after electron-ion recombination of $N_4^+$. Structure and stability of nitrogen cations, including $N_4^+$, have been analyzed in a comprehensive review [39].

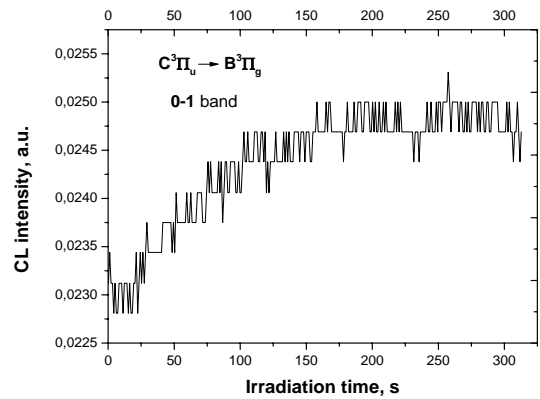

**Figure 6** Dose dependence of the second positive system (0-1 band).

To ensure contribution of charged defect centers in desorption of the excited nitrogen molecules we measured thermally stimulated luminescence TSL, that is the recombination luminescence, at different wavelengths, including wavelengths of the second positive system bands. Simultaneously the yield of TSEE was measured. Fig. 7 shows the TSEE yield as compared to the TSL yield measured in 0-1 band of the $C^3\Pi_u \rightarrow B^3\Pi_g$ system.

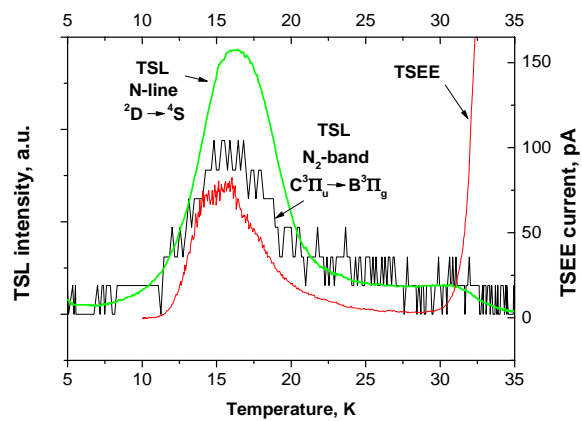

**Figure 7** Yields of the TSL measured in 0-1 band of the second positive system and α-band relative to the concurrently detected TSEE yield.

Temperature behaviour of the recombination luminescence in the $C^3\Pi_u \rightarrow B^3\Pi_g$ transition band strongly correlates with the TSEE current, indicating that the primary process, which underlies the desorption of excited nitrogen molecules, is the recombination of electron with positively charged ionic center $N_4^+$ (reaction 2). The excited molecule at the surface experiences a repulsive interaction with neighbours because of negative electron affinity of nitrogen [40] that makes effective so-called "cavity expulsion" mechanism, which is realized in light rare-gas solids [32].

The most intense feature of the TSL spectrum in visible range is the α-group, related to the $^2D \rightarrow ^4S$ atomic transition. Its observation in the TSL and correlation of the TSL yield detected in the α-band with the TSEE yield points to a connection of the atomic center with a charge recombination reaction. Recombination of $N_2^+$ with electron cannot be responsible for the TSL in α-band of pure nitrogen, while in $N_2$-doped rare-gas matrices the dissociative recombination of $N_2^+$ may result in creation of defect center $N(^2D)$ [41] by analogy with the well-known dissociative recombination in a low-density nitrogen gas [42]. Ionic center $N_4^+$ dissociates into molecules by the reaction (2). As it was mentioned $N_3^+$ centres are generated in solid nitrogen under electron bombardment [20]. The dissociative recombination of $N_3^+$ cation was investigated at the heavy-ion storage ring CRYRING [43]. Two exothermic channels were found – two-body and three-body:

$$N_3^+ + e^- \rightarrow N_2 + N + \Delta E_3 \qquad (3)$$
$$N_3^+ + e^- \rightarrow N + N + N + \Delta E_4 \qquad (4)$$

A strong propensity to dissociate through the two-body channel (3) has been found. Energy released in the channel (4), an order of magnitude smaller than that released in the channel (3), which makes it impossible to dissociate with the appearance of $N(^2D)$ by the three-body channel. $\Delta E_3$ exceeds 10 eV [43], which creates preconditions for dissociation with the emergence of $N(^2D)$ center. We observed accumulation of these centers upon irradiation as illustrated in Fig. 8.

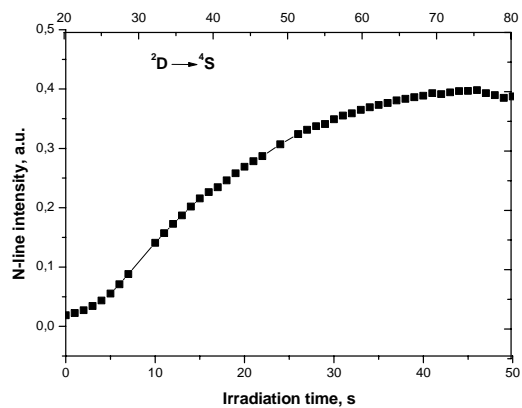

**Figure 8** Dose dependence of the α-group in luminescence.

Analysis of the data obtained counts in favor of the channel (3) as a main channel creating N($^2$D) radicals in solid N$_2$. The radiative transition $^2$D→$^4$S changes the atomic state to the ground one, leaving the atomic lattice defect.

## 3 Summary

Luminescence spectra of solid nitrogen excited by low-energy electron beam were measured over a broad range of wavelengths – from the visible region to the vacuum ultraviolet. The study of spectra evolution under irradiation provided information on defect production and accumulation, molecule fragmentation and particle desorption. Probing the samples in depth by varying electron energy permitted discrimination of the surface- and bulk-related processes. The impact of electrons is found to induce desorption of excited nitrogen molecules in the C$^3\Pi_u$ state followed by the radiative transition the C$^3\Pi_u$→B$^3\Pi_g$. The study of relaxation emission – spectrally resolved TSL and TSEE enabled us to restore the desorption scenario and elucidate the role of charged defect centers. The dissociative recombination of N$_4^+$ with electron is suggested to be a key process underlying the desorption of excited molecules. Atomic defect creation in the bulk proceeds the most likely via the reaction of dissociative recombination of N$_3^+$ with electron. These findings are in line with the detection of cluster ions (N$_2$)$_n$N$_2^+$ and (N$_2$)$_n$N$^+$ sputtering by fast ion collisions [44].

### Acknowledgements

The authors cordially thank Profs. Peter Feulner, Roman Pedrys and G. Strazzulla for stimulating discussions.

### References

[1] O. Ellegard, J. Schou, H. Sorensen, and P. Borgesen Surf. Sci. **167**, 474 (1986).
[2] R. Pedrys, D. J. Oostra, A. Haring, A. E. Devries, and J. Schou, Rad. Effects and Defects in Solids, **109**, 239 (1989).
[3] E. Hudel, E. Steinacker and P. Feulner, Surf. Sci. **273**, 405 (1992).
[4] O. Rakhovskaia, P. Wiethoff, P. Feulner, Nucl. Instr. and Meth. Phys. Res. B**101**, 169 (1995).
[5] H. Shi, P. Cloutier, L. Sanche, Phys Rev B**52**, 5385 (1995).
[6] M. Caron, H. Rothard, A. Clouvas, Surf. Sci. **528**, 103 (2003).
[7] F. L. Rook, R. E. Johnson, W. L. Brown, Surf. Sci. **164**, 625 (1985).
[8] W. L. Brown and R. E. Johnson Nucl. Instr. and Meth. Phys. Res. B**13**, 295 (1986).
[9] O. Ellegard, J. Schou, B. Stenum, H. Sorensen, R. Pedrys, B. Warczak, D. J. Oostra, A. Haring, A. E. Devries, Surf. Sci., **302**, 371 (1994).
[10] O. Ellegard, J. Schou, H. Sorensen, R. Pedrys, B. Warczak, Nucl. Instr. and Meth. Phys. Res. B**78**, 192 (1993).
[11] V. Balaji, D. E. David, R. Tian, J. Michl, H. M. Urbassek, J. Phys. Chem. **99**, 15565 (1995).
[12] B. Stenum, O. Ellegaard, J. Schou, H. Sorensen, R. Pedrys, Nucl. Instr. and Meth. Phys. Res. B**58**, 399 (1991).
[13] R. Zehr, C. French, B. C. Haynie, A. Solodukhin, I. Harrison, Surf. Sci. **451**, 76 (2000).
[14] K. I. Öberg, E. F. van Dishoeck, H. Linnartz, Astron. Astrophys. **496**, 281 (2009).

[15] R. E. Johnson, R. W. Carlson, T. A. Cassidy, and M. Fama, in: The Science of Solar System Ices, edited by M. S. Gudipati and J. Castillo-Rogez, Astrophysics and Space Science Library, Vol. 356 (Springer Science and Business Media, New York, 2012), chap. 17.
[16] I. V. Khyzhniy, E. V. Savchenko, S. A. Uyutnov, G. B. Gumenchuk, A. N. Ponomaryov, V. E. Bondybey, Radiation Measurements, **45**, 353 (2010).
[17] V. G. Storchak, D. G. Eshchenko, J. H. Brewer, S. P. Cottrell, S. F. J. Cox, E. Karlsson, R. W. Wappling, J. Low Temp. Phys. **122**, 527 (2001).
[18] R. E. Johnson and J. Schou, Mat. Fys. Medd. Dan. Vid. Selsk. **43**, 403 (1993).
[19] E. V. Savchenko, I. V. Khyzhniy, S. A. Uyutnov, A. N. Ponomaryov, G. B. Gumenchuk and V. E. Bondybey, Low Temp. Phys. **39**, 446 (2013).
[20] Y-J. Wu, H-F. Chen, S-J. Chuang, T-P. Huang, Astrophys. J. **768**, 83 (2013).
[21] M. A. Allodi, R. A. Baragiola, G. A. Baratta, M. A. Barucci, G. A. Blake, J. R. Brucato, C. Contreras, S. H. Cuylle, Ph. Boduch, D. Fulvio, M. S. Gudipati,,S. Ioppolo, Z. Kaňuchová, A. Lignell, H. Linnartz, M. E.Palumbo, U. Raut, H. Rothard, F. Salama, E. V. Savchenko, E. Sciamma-O'Brien, G. Strazzulla, Space Sci. Rev. **180**, 101 (2013).
[22] E.V. Savchenko and V. E. Bondybey, Phys. Stat. Sol. (a) **202**, 221 (2005).
[23] O. Oehler, D. A. Smith, and K. Dressler, J. Chem. Phys. **66**, 2097 (1977).
[24] F. Coletti and A. M. Bonnot, Chem. Phys. Lett. **45**, 580 (1977).
[25] Yu. B. Poltoratskii, V. M. Stepanenko, and I. Ya. Fugol', Low Temp. Phys. **7**, 60 (1981).
[26] I. Ya. Fugol', Yu. B. Poltoratski, and Yu. I. Rybalko, Low Temp. Phys. **4**, 496 (1978).
[27] R. S. Mangina, J. M. Ajello, R. A. West, and D. Dziczek, ApJS **196**, 13 (2011).
[28] S. Zuochun, L. Jianye, A. H. Hamdani, G. Huide. And M. Zuguang, Sci. Chine. **46**, 89 (2003).
[29] A. Adams and P. K. Hansma, Phys. Rev. B **22**, 4258 (1980).
[30] V. A. Slusarev, Yu. A. Freiman, I. N. Krupskii, I. A. Burakhovich, Phys. Status Solidi B, **54**, 745 (1972).
[31] D. S. Tinti, G. W. Robinson, J. Chem. Phys. **49**, 3229 (1968).
[32] K. S. Song, and R. T. Williams, Self-Trapped Excitons (Springer-Verlag, Berlin Heidelberg New York, 1996), p. 404.
[33] A. M. Boichenko, V. A. Tarasenko, and S. I. Yakovlenko, Laser Phys. **9**, 1004 (1999).
[34] K. Carleton, K. Welge, and S. Leone, Chem. Phys. Lett. **115**, 492 (1985).
[35] I. K. Randeniya, X. K. Zeng. R. S. Smith, and M. A. Smith, J. Phys. Chem. **93**, 8031 (1989).
[36] Y. S. Cao and R. Johnsen, J. Chem. Phys. **95**, 7356 (1991).
[37] M. C Sauer, Jr., and W. A. Mulac, J. Chem. Phys. **56**, 4995 (1972).
[38] S. F. Adams, C. A. Dr Joseph, Jr., and J. M. Williamson, J. Chem. Phys. **130**, 144316 (2009).
[39] M. T. Nguyen, Coord. Chem. Rev. **244**, 93 (2003).
[40] G. Bader, G. Perluzzo, L. G. Caron, and L. Sanche, Phys. Rev. B **30**, 78 (1984).
[41] E. V. Savchenko, I. V. Khyzhniy, S. A. Uyutnov, G. B. Gumenchuk, A. N. Ponomaryov, and V. E. Bondybey, IOP Conf. Series: Materials Science and Engineering, **15**, 012082 (2010).
[42] J. L. Fox and A. Dalgarno, J. Geophys. Res. **88**, 9027 (1983).
[43] V. Zhaunerchyk, W. D. Geppert, E. Vigren, M. Hamberg, M. Danielsson, m. Larsson, R. D. Thomas, M. Kaminska, F. Österdahl, J. Chem. Phys. **127**, 014305 (2007).
[44] L. S. Farenzena, P. Iza, R. Martinez, F. A. Fernandez-Lima, E. Seperuelo Duarte, G. S. Faraudo, C. R. Ponciano, E. F. da Silveira, M. G. P. Homem, A. Naves de Brito, K. Wien, Earth Moon Planets **97**, 311 (2005).